\documentclass[aps,preprint]{revtex4}%
\usepackage{amsfonts}
\usepackage{amsmath}
\usepackage{amssymb}
\usepackage{graphicx}%
\begin{document}
\preprint{hep-th/0506045}
\title[ ]{A solution to Ward-Takahasi-identity in QED$_{3}$}
\author{Yuichi Hoshino}
\affiliation{Kushiro National College of Technology,Otanoshike Nishi -32-1,Kushiro,Hokkaido,084-0916,Japan}
\author{Second Author}
\affiliation{My Institution}
\author{Third Author}
\affiliation{Other Institution}

\begin{abstract}
Using spectral function of photon we find the reliable results for the effects
of vacuum polarization for the dressed fermion propagator in three-dimensional QED.

\end{abstract}
\volumeyear{year}
\volumenumber{number}
\issuenumber{number}
\eid{identifier}
\date[Date text]{date}
\received[Received text]{date}

\revised[Revised text]{date}

\accepted[Accepted text]{date}

\published[Published text]{date}

\startpage{101}
\endpage{102}
\maketitle
\tableofcontents

\section{\qquad Introduction}

In the previous work we studied the mass singularity of the fermion propagator
in QED$_{3}$ in the presence of massless photon in quenched
approximation[7].We applied the method of spectral function with low-energy
theorem which is known to reproduces the Bloch-Nordsieck approximation and
renormalization group analysis near the mass shell in four-dimensional
model[5,6,12].In the present work we study the effect of vacuum polarization
of massless fermion on the photon propagator in the same approximation.Using
the spectral function of photon we get the non-perturbative effects by
integrating the quenched case with bare photon mass.However high energy
behaviour of the fermion propagator does not change and we get $Z_{2}^{-1}=0$
for arbitrary coupling.At sufficently large coupling short distance
singularities disappear.In this case the vacuum expectation value
$\left\langle \overline{\psi}\psi\right\rangle $ becomes finite.In Section II
we rewiew the vacuum polarization of the massive fermion loop and show the
structure of the photon propagator.In section III spectral function of the
fermion propagator are defined and we show the way to determine it based on
LSZ reduction formula and low-energy theorem.In section IV we evaluate the
spectral function for quenched case with photon mass as an infrarfed cut-off
and improve it by the photon spectral function including vacuum
polarization.In Section V is devoted to analysis in momentum space of these
solutions.In section VI we consider the origins of confinement in the
modification of Coulomb potential by LSZ.

\section{Vacuum polarization}

Assuming parity invariance,we take 4-component fermion
representation[6,8,9,10].The one-loop self-energy for photon with dimesional
regularization is given
\begin{align}
\Pi_{\mu\nu}(k)  &  \equiv ie^{2}\int\overline{d}^{3}ptr(\gamma_{\mu}\frac
{1}{\gamma\cdot p-m}\gamma_{\nu}\frac{1}{\gamma\cdot(p-k)-m})\nonumber\\
&  =-e^{2}\frac{T_{\mu\nu}}{8\pi}[(\sqrt{k^{2}}+\frac{4m^{2}}{\sqrt{k^{2}}%
})\ln(\frac{2m+\sqrt{k^{2}}}{2m-\sqrt{k^{2}}})-4m],\nonumber\\
T_{\mu\nu}  &  =-(g_{\mu\nu}-\frac{k_{\mu}k_{\nu}}{k^{2}}),\overline{d}%
^{3}p=\frac{d^{3}p}{(2\pi)^{3}}.
\end{align}
Usual polarization function $P(k)$ is expanded in terms of $m(1/m)$%
\begin{align}
P(k)  &  =-\frac{e^{2}}{8\pi}[(\sqrt{-k^{2}}+\frac{4m^{2}}{\sqrt{-k^{2}}}%
)\ln(\frac{2m+\sqrt{-k^{2}}}{2m-\sqrt{-k^{2}}})-4m],\nonumber\\
&  =+\frac{e^{2}}{8}\sqrt{-k^{2}}-4\frac{m^{2}}{\sqrt{-k^{2}}}+\frac{64m^{3}%
}{3\pi(-k^{2})}-\frac{512m^{5}}{15\pi(-k^{2})^{2}}+O(m^{6}))(k^{2}%
<0),\nonumber\\
&  =+\frac{e^{2}}{6\pi m}(-k^{2})-\frac{e^{2}}{60\pi m^{3}}(-k^{2}%
)^{2}+O(k^{5}).
\end{align}
If the mass $m$ is heavy expansion in terms of the inverse powers of $m$ is a
weak coupling expansion and massless limit corresponds to a strong coupling
limit.Here we see that the massless limit is a correct high-energy behaviour
of the vacuum polarization function $P.$The full propagator is
\begin{equation}
D_{F}(k)=-\frac{T_{\mu\nu}}{k^{2}-P(k)}-d\frac{k_{\mu}k_{\nu}}{k^{4}}.
\end{equation}
To see the screening effects easily,first we set $m=0$ to neglect the
threshold effects.It is equivalent to study the high-energy behaviour in
momentum space.Hereafter we use the full photon propagator with $N$ fermion
flavours%
\begin{equation}
D_{F}(k)=-\frac{T_{\mu\nu}}{k^{2}+e^{2}N/8\sqrt{k^{2}}}+g.f.
\end{equation}
and do not discuss the analyticity in Minkowski space.

\section{Calculating the spectraly weighted propagator}

\subsection{\bigskip Definition of the spectral function}

In this section we show how to evalute the fermion propagator non
pertubatively by the spectral represntation[5,7].The spectral function of the
fermion is defined
\begin{align}
S_{F}(s^{\prime})  &  =P\int ds\frac{\gamma\cdot p\rho_{1}(s)+\rho_{2}%
(s)}{s^{\prime}-s}+i\pi(\gamma\cdot p\rho_{1}(s^{\prime})+\rho_{2}(s^{\prime
})),\nonumber\\
\rho(p)  &  =\frac{1}{\pi}\operatorname{Im}S_{F}(p)=\gamma\cdot p\rho
_{1}(p)+\rho_{2}(p)\\
&  =(2\pi)^{2}\sum_{n}\delta(p-p_{n})\left\langle 0|\psi(x)|n\right\rangle
\left\langle n|\overline{\psi}(0)|0\right\rangle .
\end{align}
In the quenched approximation the state $|n>$ stands for a fermion and
arbitrary numbers of photons,%
\begin{equation}
|n>=|r;k_{1},...,k_{n}>,r^{2}=m^{2},
\end{equation}
we have the solution which is written symbolically%
\begin{align}
\rho(p)  &  =\int\frac{md^{2}r}{r^{0}}\sum_{n=0}^{\infty}\frac{1}{n!}\left(
\int\overline{d^{3}}k\theta(k^{0})\delta(k^{2})\sum_{\epsilon}\right)
_{n}\delta(p-r-\sum_{i=1}^{n}k_{i})\nonumber\\
&  \times\left\langle \Omega|\psi(x)|r;k_{1},...,k_{n}\right\rangle
\left\langle r;k_{1},...,k_{n}|\overline{\psi}(0)|\Omega\right\rangle .
\end{align}
Here the notations
\[
(f(k))_{0}=1,(f(k))_{n}=%
{\displaystyle\prod\limits_{i=1}^{n}}
f(k_{i})
\]
have been introduced to show the phase space of each photons.We consider the
matrix element
\begin{equation}
T_{n}=\left\langle \Omega|\psi|r;k_{1},....,k_{n}\right\rangle
\end{equation}
for $k_{n}^{2}\neq0$ by LSZ reduction formula:%
\begin{align}
T_{n}  &  =\epsilon_{n}^{\mu}T_{n\mu},\nonumber\\
\epsilon_{n}^{\mu}T_{n}^{\mu}  &  =\frac{i}{\sqrt{Z_{3}}}\int d^{3}%
y\exp(ik_{n}\cdot y)\square_{y}\left\langle \Omega|T\psi(x)\epsilon^{\mu
}A_{\mu}(y)|r;k_{1},...,k_{n-1}\right\rangle \nonumber\\
&  =-\frac{i}{\sqrt{Z_{3}}}\int d^{3}y\exp(ik_{n}\cdot y)\left\langle
\Omega|T\psi(x)\epsilon^{\mu}j_{\mu}(y)|r;k_{1},...,k_{n-1}\right\rangle ,
\end{align}
provided%
\begin{align}
\square_{x}T(\psi A_{\mu}(x)  &  =T\psi\square_{x}A_{\mu}(x)=T\psi(-j_{\mu
}(x)+\frac{d-1}{d}\partial_{\mu}^{x}(\partial\cdot A(x)),\\
\partial\cdot A^{(+)}|phys  &  >=0,
\end{align}
where $d$ is a gauge fixing parameter.$T_{n}$ satisfies
Ward-Takahashi-identity%
\begin{equation}
k_{n\mu}T_{\mu}^{n}(r;k_{1},...,k_{n})=e\exp(ik_{n}\cdot x)\left\langle
\Omega|\psi(x)|r;k_{1},....,k_{n-1}\right\rangle ,
\end{equation}
provided%
\begin{equation}
\partial_{\mu}^{y}T(\psi(x)J_{\mu}(y))=-e\psi(x)\delta(x-y).
\end{equation}
Neglecting position dependence which is given by%
\begin{equation}
\psi(x)=\exp(-ip\cdot x)\psi(0)\exp(ip\cdot x),
\end{equation}
we get the usual form%
\begin{equation}
k_{n\mu}T_{\mu}^{n}(r;k_{1},...k_{n-1}),r^{2}=m^{2},
\end{equation}
which implies low-energy theorems.By the low-energy theorem the fermion pole
term for the external line is dominant for the infrared singularity for
fermion.Let us consider the matrix element $T_{n-1}(r;k_{1},...k_{n-1})$ in
which there is no photon line with external fermion.We attach $T_{n-1}$ with
one photon line with external fermion.The matrix element $T_{n}$ is written
as
\begin{equation}
T_{n}^{(pole)}(r;k_{1},....k_{n})=T_{n-1}(r+k_{n};k_{1},...k_{n-1}%
)T_{1}(r;k_{n}),
\end{equation}
here we assume $T_{n-1}(r+k_{n};k_{1},...k_{n-1})$ is regular at $k_{n}%
^{2}=0.$Of course only $T^{(pole)}$ does not satisfy
Ward-Takahashi-identity.In ref [5],the method to determine the matrix element
$T_{n}$ was discussed in the following way up to $k_{\mu}^{n}$ terms which
vanishes in the limit $k_{n}^{\mu}=0,$
\begin{align}
T_{n}^{\mu}  &  =T_{n}^{\mu(pole)}+T_{n}^{\mu\prime}+R_{n}^{\mu},\\
k_{n\mu}T_{n}^{\mu^{\prime}}  &  =eT_{n-1}-k_{n\mu}T_{n}^{\mu(pole)},\\
k_{n\mu}R_{n}^{\mu}  &  =0.
\end{align}
The result for $T_{n}^{\mu(pole)}$ is
\begin{equation}
T_{n}^{\mu(pole)}=T_{1}(r+k_{n};k_{n})\Lambda_{n-1}(r+k_{n};k_{1}%
,...,k_{n-1}).
\end{equation}
Here,$\Lambda_{n-1}(r;k_{1},...,k_{n-1})$ does not contain any one-fermion
line with momentum $r+k_{n}$(off-shell) and coincides with $T_{n-1}%
(r;k_{1},..,k_{n-1})$ continued off the $r^{2}$ mass shell(by an LSZ formula
for example):%
\begin{equation}
T_{n-1}(r;k_{1},...,k_{n-1})=\Lambda_{n-1}(r;k_{1},...,k_{n-1})|_{r^{2}=m^{2}%
}.
\end{equation}
For $n=1$ case we have
\begin{equation}
k_{\mu}T_{1}(r+k;k)=k_{\mu}\frac{e\gamma^{\mu}}{\gamma\cdot(r+k)-m}%
U(r)=e\left\langle \Omega|\psi|r\right\rangle =eU(r).
\end{equation}
From (17),(19),(23) we see that $T_{n}^{\mu^{\prime}}$ is sufficent to satisfy%
\begin{align}
k_{n\mu}T_{n}^{\mu^{\prime}}(r;k_{1},..k_{n})  &  =e[T_{n-1}(r;k_{1}%
,...,k_{n-1})-\Lambda_{n-1}(r+k_{n};k_{1},..,k_{n-1})]\nonumber\\
&  =-ek_{n}^{\mu}\frac{\partial}{\partial k_{n}^{\mu}}[\Lambda_{n-1}%
(r+k_{n};k_{1},..,k_{n-1})]_{k_{n}=k_{n}^{\ast}},
\end{align}
where $k^{\ast}$ is a intermediate point for $k$.From the above it is
sufficent to take
\begin{equation}
T_{n}^{\mu^{\prime}}=-e\frac{\partial}{\partial k_{n\mu}}\Lambda_{n-1}%
(r+k_{n};k_{1},..,k_{n-1})|_{k_{n}=k_{n}^{\ast}}=-e\frac{\partial}{\partial
r_{\mu}^{^{\prime}}}\Lambda_{n-1}(r^{\prime};k_{1},..,k_{n-1})|_{r^{^{\prime}%
}=r+k_{n^{\ast}}.}%
\end{equation}%
\begin{equation}
T_{n}^{\mu(pole)}+T_{n}^{\mu^{\prime}}=\frac{e\gamma^{\mu}}{\gamma
\cdot(r+k_{n})-m}T_{n-1}+e(\frac{\gamma^{\mu}k_{n}^{\nu}}{\gamma\cdot
(r+k_{n})-m}-g^{\mu\nu})\frac{\partial}{\partial r^{^{\prime}\nu}}%
\Lambda_{n-1}|_{r^{\prime}=r+k^{\ast},r^{\prime2}=m^{2}+\epsilon}.
\end{equation}
The second term is transverse in $k_{n}^{\mu}$ and should be regular at
$k_{n}^{\mu}=0.$It is understood as the non leading term and derived by gauge invariance.

\subsection{\bigskip Approximation to the spectral function}

If there are massless particle as photons,there exists infrared divergences
near the fermion mass-shell.In theses cases we cannot separate the one
particle state and one particle with multiphoton intermediate states,we must
sum all intermediate states with infinite numbers of photons.The spectral
representation of the propagator in position space is given.%
\begin{equation}
S_{F}(x)=\int\overline{d}^{3}p\exp(p\cdot x)\int\frac{ds\rho(s)}{p^{2}+s}.
\end{equation}
First we derive the second-order spectral function and the propagator
$S_{F}(p)$%
\begin{equation}
\overline{\rho}(x)=-e^{2}\int\frac{md^{2}r}{(2\pi)^{2}r^{0}}\exp(ir\cdot
x)\int\overline{d}^{3}k\theta(k^{0})\delta(k^{2})\exp(ik\cdot x)\sum
_{\lambda,s}T_{1}\overline{T}_{1},
\end{equation}
where one-photon matrix element $T_{1}$ which is given in ref[7,10]%
\begin{align}
&  T_{1}=\left\langle \Omega|\psi(x)|r;k\right\rangle =\left\langle
in|T(\psi_{in}(x),ie\int d^{3}y\overline{\psi}_{in}(y)\gamma_{\mu}\psi
_{in}(y)A_{in}^{\mu}(y)|r;k\text{ }in\right\rangle \nonumber\\
&  =ie\int d^{3}yd^{3}zS_{F}(x-z)\gamma_{\mu}\delta^{(3)}(y-z)\exp(i(k\cdot
y+r\cdot z))\epsilon^{\mu}(k,\lambda)U(r,s)\nonumber\\
&  =-ie\frac{1}{(r+k)\cdot\gamma-m+i\epsilon}\gamma_{\mu}\epsilon^{\mu
}(k,\lambda)\exp(i(r+k)\cdot x)U(r,s).
\end{align}%
\begin{align}
\sum_{\lambda,s}T_{1}\overline{T}_{1}  &  =\frac{(r+k)\cdot\gamma+m}%
{(r+k)^{2}-m^{2}}\gamma^{\mu}\frac{(\gamma\cdot r+m)}{2m}\frac{(r+k)\cdot
\gamma+m}{(r+k)^{2}-m^{2}}\gamma^{\nu}\Pi_{\mu\nu},\nonumber\\
r^{2}  &  =m^{2},k^{2}=0
\end{align}
and $\delta(k^{2})$ is a bare photon spectral function.Here $\Pi_{\mu\nu}$ is
the polarization sum%
\begin{equation}
\Pi_{\mu\nu}=\sum_{\lambda}\epsilon_{\mu}(k,\lambda)\epsilon_{\nu}%
(k,\lambda)=-g_{\mu\nu}-(d-1)\frac{k_{\mu}k_{\nu}}{k^{2}}.
\end{equation}
In this case it is easy to find the explicit form of $\rho$%
\begin{equation}
\rho^{(2)}(p)=\int d^{3}x\exp(-ip\cdot x)\overline{\rho}(x),
\end{equation}%
\begin{align}
\rho^{(2)}(p)=-e^{2}\int\frac{d^{2}r}{(2\pi)^{2}r^{0}}  &  \int\frac{d^{2}%
k}{(2\pi)^{2}2k^{0}}\delta^{(3)}(p-k-r)\nonumber\\
&  \times\frac{1}{2}(\gamma\cdot r+m)[[\frac{m^{2}}{(r\cdot k)^{2}}+\frac
{1}{r\cdot k}]+\frac{d-1}{k^{2}}].
\end{align}
At $d=1$ gauge,$\rho$ is evaluated in the center of mass system
\begin{align}
(2\pi)^{2}\rho^{(2)}(p)  &  =\frac{\pi e^{2}}{2\sqrt{p^{2}}}\sum_{\lambda
,s}T_{1}\overline{T}_{1}\theta(p^{2}-m^{2})\nonumber\\
&  =\frac{\pi e^{2}}{2\sqrt{p^{2}}}(\gamma\cdot p+m)[\frac{4m^{2}}%
{(p^{2}-m^{2})^{2}}+\frac{2}{p^{2}-m^{2}}]\theta(p^{2}-m^{2}).
\end{align}
In this order we have the propagator in specrtral form%
\begin{align}
S^{(2)}(p)  &  =\int_{m^{2}}^{\infty}ds\frac{\rho^{(2)}(s)}{p^{2}-s+i\epsilon
}\nonumber\\
&  =\frac{e^{2}}{4\pi}\int_{m^{2}}^{\infty}ds\frac{(\gamma\cdot p+m)}%
{p^{2}-s+i\epsilon}\frac{1}{s}[\frac{4m^{2}}{(s-m^{2})^{2}}+\frac{2}{s-m^{2}%
}].
\end{align}
From the above equation we see that $S^{(2)}$ has infrared divergences near
$p^{2}=m^{2}.$If we sum infinite number of photons in the final state as in
(8),by LSZ reduction formula the method to determine the multi-photon matrix
element is given in (26).In the lowest order approximation simplest solution
to the spectral function is given by exponentiation of one photon matrix
element $T_{1}\overline{T_{1}}$:%
\begin{equation}
\left\langle \Omega|\psi(x)|r;k_{1},...,k_{n}\right\rangle \left\langle
r;k_{1},..k_{n},\overline{\psi}(0)|\Omega\right\rangle \rightarrow%
{\displaystyle\prod\limits_{j=1}^{n}}
T_{1}(k_{j})\overline{T}_{1}(k_{j}),
\end{equation}%
\[
\overline{\rho}(x)=\int\frac{md^{2}r}{(2\pi)^{2}r^{0}}\exp(ir\cdot x)\exp(F),
\]%
\begin{align}
F  &  =\sum_{one\text{ }photon}\left\langle \Omega|\psi(x)|r;k\right\rangle
\left\langle r;k|\overline{\psi}(0)|\Omega\right\rangle \nonumber\\
&  =\int\overline{d^{3}}k\delta(k^{2})\theta(k^{0})\exp(ik\cdot x)\sum
_{\lambda,s}T_{1}\overline{T_{1}}.
\end{align}
This approximation leads to an infinite sum of ladder graphs with fixed
mass.And its imaginary part is simple to evaluate by the above formula.After
integration,we set $r^{2}=m^{2}$%
\begin{align}
S_{F}(x)  &  =\int\overline{d}^{3}p\exp(p\cdot x)(i\gamma\cdot\partial
+m)\frac{\exp(-m\left\vert x\right\vert )}{2\times4\pi\left\vert x\right\vert
}\nonumber\\
&  \times\exp(-e^{2}\int\overline{d}^{3}k\exp(ik\cdot x)\theta(k^{0}%
)\delta(k^{2})[\frac{m^{2}}{(r\cdot k)^{2}}+\frac{1}{(r\cdot k)}+\frac
{d-1}{k^{2}}]).
\end{align}
This formula shows that the infrared divergences of photons give the radiative
correction to the free position space propagator $\exp(-m\left\vert
x\right\vert )/\left\vert x\right\vert $.The radiative correction for the
fermion propagator may be seen as the modification of the short distance
behaviour or mass.Therefore we expand $F$
\begin{equation}
S(x)=\frac{\exp(-m\left\vert x\right\vert )}{4\pi\left\vert x\right\vert }%
\exp(F(m,x))
\end{equation}
in the power series of $x$ and study the correction of mass.The results are
shown in the previous work.Next we use the full propagator for photon in the
same way%
\begin{align}
D_{F}(x)  &  =\int\int\overline{d}^{3}k\exp(ik\cdot x)\frac{\rho(s)ds}%
{k^{2}-s+i\epsilon}\nonumber\\
&  =\int ds\int\overline{d}^{3}k\exp(ik\cdot x)[P\frac{\rho(s)}{k^{2}-s}%
-i\pi\rho(s)\delta(k^{2}-s)]\nonumber\\
&  =\int\overline{d}^{3}k\exp(ik\cdot x)[\operatorname{Re}D_{F}%
(k)+i\operatorname{Im}D_{F}(k)].
\end{align}
Spectral functions for free and dressed photon are given by [6,11]%
\begin{align}
\rho^{(0)}(k)  &  =\delta(k^{2}-\mu^{2}),\nonumber\\
\rho^{D}(k)  &  =\frac{1}{\pi}\operatorname{Im}D_{F}(k)=\frac{c\sqrt{k^{2}}%
}{k^{2}(k^{2}+c^{2})}.
\end{align}
One photon matrix element reads%
\begin{equation}
2F=-e^{2}\int\overline{d}^{3}k\exp(ik\cdot x)[i\operatorname{Im}D_{F}%
(k)[\frac{m^{2}}{(r\cdot k)^{2}}+\frac{1}{(r\cdot k)}-\frac{1}{k^{2}}%
]+\frac{d}{k^{4}}].
\end{equation}
\qquad\qquad\ First we perform the direct fourier transform of the propagator%
\begin{align}
iD_{F}(x)  &  =\int\overline{d}^{3}k\exp(ik\cdot x)\frac{1}{k^{2}+c\sqrt
{k^{2}}}\nonumber\\
&  =\frac{1}{2\pi^{2}\left\vert x\right\vert }\int_{0}^{\infty}dk\frac
{\sin(\sqrt{k^{2}}\left\vert x\right\vert )}{\sqrt{k^{2}}+c},c=\frac{e^{2}%
N}{8}.
\end{align}
On the other hand the spectral representation for photon is%
\begin{align}
iD_{F}(x)  &  =\frac{1}{\pi}\int\overline{d}^{3}k\exp(ik\cdot x)\int
_{0}^{\infty}dp^{2}\frac{\rho(p)}{k^{2}+p^{2}}\nonumber\\
&  =\int_{0}^{\infty}dp^{2}\frac{\exp(-\sqrt{p^{2}}\left\vert x\right\vert
)}{4\pi\left\vert x\right\vert }\rho(p^{2}).
\end{align}
Both ways lead the same answer;%
\begin{equation}
iD_{F}(x)=\frac{1}{4\pi^{2}\left\vert x\right\vert }[\pi\cos(c\left\vert
x\right\vert )-2\operatorname{Si}(c\left\vert x\right\vert )\cos(c\left\vert
x\right\vert )+2\operatorname{Ci}(c\left\vert x\right\vert )\sin(c\left\vert
x\right\vert )].
\end{equation}
These modification leads to the change of the static potential

\qquad%
\begin{align}
V_{B}(x)  &  =\int\overline{d}^{2}k\exp(ik\cdot x)\frac{1}{k^{2}+\mu^{2}%
}=K_{0}(\mu\left\vert x\right\vert ),\\
V_{R}(x)  &  =\int\overline{d}^{2}k\exp(ik\cdot x)\frac{1}{k^{2}+ck}%
=\frac{H_{0}(c\left\vert x\right\vert )-Y_{0}(c\left\vert x\right\vert )}%
{8},\\
H_{0}(a)  &  =\frac{2}{\pi}\int_{0}^{1}\frac{\sin(ax)}{\sqrt{1-x^{2}}}%
dx,Y_{0}(a)=-\frac{2}{\pi}\int_{1}^{\infty}\frac{\cos(ax)}{\sqrt{x^{2}-1}}dx.
\end{align}
$V_{R}(x)$ has not zero on the $\left\vert x\right\vert $ axis and decrease as
$1/c\left\vert x\right\vert $.On the other hand $V_{B}(x)$ has zero and change
its sign$.$

\section{Approximate solution in position space}

\subsection{quenched case}

To evaluate the the function $F$,it is helpful to use the following parameter
integral with exponential cut-off(infrared cut-off)[5,7,12].Following the
parameter tric%
\begin{align}
\frac{1}{k\cdot r}  &  =i\lim_{\epsilon\rightarrow0}\int_{0}^{\infty}%
d\alpha\exp(i\alpha(k+i\epsilon)\cdot r),\nonumber\\
\frac{1}{(k\cdot r)^{2}}  &  =-\lim_{\epsilon\rightarrow0}\int_{0}^{\infty
}\alpha d\alpha\exp(i\alpha(k+i\epsilon)\cdot r),
\end{align}
we obtain the formulea
\begin{equation}
F_{1}=\int\overline{d}^{3}k\exp(ik\cdot x)D_{F}(k)\frac{1}{(k\cdot r)^{2}%
}=-\lim_{\mu\rightarrow0}\int_{0}^{\infty}\alpha d\alpha D_{F}(x+\alpha
r)\exp(-\mu\alpha r).
\end{equation}%
\begin{equation}
F_{2}=\int\overline{d}^{3}k\exp(ik\cdot x)D_{F}(k)\frac{1}{k\cdot r}%
=i\lim_{\mu\rightarrow0}\int_{0}^{\infty}d\alpha D_{F}(x+\alpha r)\exp
(-\mu\alpha r).
\end{equation}
Soft photon divergence corresponds to the large $\alpha$ region and $\mu$ is
an infrared cut-off.For the $k_{\mu}k_{\nu}$ part there remains an infrared
divergence $1/k^{2}$ which is independent of the $1/(r\cdot k)$ in the same
gauge in (42)$.$It is simple to evaluate this term $F_{L}$ by definition%

\begin{equation}
F_{L}=-ie^{2}\int\overline{d}^{3}k\exp(ik\cdot x)D_{F}(k)\frac{1}{k^{2}},
\end{equation}%
\begin{equation}
\frac{1}{4\pi^{2}}\int_{0}^{\infty}d\sqrt{k^{2}}\frac{\sin(\sqrt{k^{2}%
}\left\vert x\right\vert )}{\sqrt{k^{2}}\left\vert x\right\vert (k^{2}+\mu
^{2})}=\frac{1-\exp(-\mu\left\vert x\right\vert )}{8\pi\mu^{2}\left\vert
x\right\vert }.
\end{equation}
We have%
\begin{equation}
F=ie^{2}m^{2}\int_{0}^{\infty}\alpha d\alpha D_{F}(x+\alpha r)-e^{2}\int
_{0}^{\infty}d\alpha D_{F}(x+\alpha r)+ie^{2}\int\overline{d}^{3}k\exp(ik\cdot
x)D_{F}(k)\frac{1}{k^{2}}.
\end{equation}
Here we notice that the over all sign is changed by replacing the imaginary
part to real part of the photon propagator.In quenced case the above formulea
for the evaluation of three terms in $F$ provided the position space
propagator with bare mass
\begin{equation}
D_{F}^{(0)}(x)_{+}=\frac{\exp(-\mu\left\vert x\right\vert )}{8\pi i\left\vert
x\right\vert }.
\end{equation}

\begin{equation}
F=-\frac{e^{2}}{8\pi}(\frac{\exp(-\mu\left\vert x\right\vert )}{\mu
}-\left\vert x\right\vert \operatorname{Ei}(\mu\left\vert x\right\vert
))-\frac{e^{2}}{8\pi m}\operatorname{Ei}(\mu\left\vert x\right\vert
)+(d-1)\frac{e^{2}}{8\pi\mu^{2}\left\vert x\right\vert }(1-\exp(-\mu\left\vert
x\right\vert )),
\end{equation}
where
\begin{equation}
\operatorname{Ei}(z)=\int_{1}^{\infty}\frac{\exp(-zt)}{t}dt.
\end{equation}
For the leading order in $\mu$ we obtain%
\begin{align}
F_{1}  &  =\frac{e^{2}}{8\pi}(-\frac{1}{\mu}+\left\vert x\right\vert
(1-\ln(\mu\left\vert x\right\vert -\gamma))+O(\mu),\\
F_{2}  &  =\frac{e^{2}}{8\pi m}(\ln(\mu\left\vert x\right\vert )+\gamma
)+O(\mu),\\
F_{g}  &  =\frac{e^{2}}{8\pi}(\frac{1}{\mu}-\frac{\left\vert x\right\vert }%
{2})(d-1)+O(\mu).
\end{align}%
\begin{equation}
F=\frac{e^{2}(d-2)}{8\pi\mu}+\frac{\gamma e^{2}}{8\pi m}+\frac{e^{2}}{8\pi
m}\ln(\mu\left\vert x\right\vert )-\frac{e^{2}}{8\pi}\left\vert x\right\vert
\ln(\mu\left\vert x\right\vert )-\frac{e^{2}}{16\pi}\left\vert x\right\vert
(d-3+2\gamma),
\end{equation}
where $\gamma$ is an Euler constant.In this case linear infrared divergence
may cancells by higher order correction or away from threshold,at present we
omitt them here with constant term[6,13].Linear term in $\left\vert
x\right\vert $ is understood as the finite mass shift from the form of the
propagator in position space and $\left\vert x\right\vert \ln(\mu\left\vert
x\right\vert )$ term is position dependent mass
\begin{align}
m  &  =\left\vert m_{0}+\frac{e^{2}}{16\pi}(d-3+2\gamma)\right\vert
,\nonumber\\
m(x)  &  =m+\frac{e^{2}}{8\pi}\ln(\mu\left\vert x\right\vert ),
\end{align}
which we will discuss in section VI.The position space propagator is written
as free one multiplied by quantum correction as%
\begin{align}
\frac{\exp(-m\left\vert x\right\vert )}{4\pi\left\vert x\right\vert }\exp(F)
&  =\frac{\exp(-m\left\vert x\right\vert )}{4\pi\left\vert x\right\vert }%
(\mu\left\vert x\right\vert )^{D-C\left\vert x\right\vert },\nonumber\\
D  &  =\frac{e^{2}}{8\pi m},C=\frac{e^{2}}{8\pi}.
\end{align}
From the above form we see that $D$ acts to change the power of $\left\vert
x\right\vert $ and plays the role of anomalous dimension of the
propagator[7].If $D\geq1$ there is no short distance singularities as spike.

\subsection{\bigskip unquenched case}

Here we apply the spectral function of photon to evaluate the unquenched
fermion proagator.We simply integrate the function $F(x,\mu)$ for quenched
case which is given in (61),where $\mu$ is a photon mass.Spectral function of
photon is given in (41) in the Landau gauge%
\begin{align}
\rho(\mu) &  =\frac{c}{\mu(\mu^{2}+c^{2})},\nonumber\\
Z_{3}^{-1} &  =\int_{0}^{\infty}\rho(\mu)\mu d\mu=\frac{\pi}{2}.
\end{align}
An improved $F$ is written as dispersion integral%
\begin{align}
\widetilde{F} &  =Z_{3}\int_{0}^{\infty}F(\mu)\rho(\mu)\mu d\mu\nonumber\\
&  =\frac{e^{2}}{8c}(-2)\ln(\frac{\mu}{c})+\frac{\gamma e^{2}}{8\pi m}%
+\frac{e^{2}}{8\pi m}\ln(c\left\vert x\right\vert )\nonumber\\
&  -\frac{e^{2}}{8\pi}\left\vert x\right\vert \ln(c\left\vert x\right\vert
)-\frac{e^{2}}{16\pi}\left\vert x\right\vert (3-2\gamma).
\end{align}
In this way the linear infrared divergences turn out to be a logarithmic
divergence in the first term.This is the improvement by spectral function.The
fermion spectral function in position space becomes%
\begin{align}
\overline{\rho}(x) &  =\frac{\exp(-m\left\vert x\right\vert )}{4\pi\left\vert
x\right\vert }\exp(\widetilde{F})\nonumber\\
&  =\frac{\exp(-(m+B)\left\vert x\right\vert )}{4\pi\left\vert x\right\vert
}(c\left\vert x\right\vert )^{D-C\left\vert x\right\vert }(\frac{\mu}%
{c})^{\beta}\exp(\frac{\gamma e^{2}}{8\pi m}),
\end{align}
by $N$ flavours%
\begin{equation}
B=\frac{c}{2N\pi}(3-2\gamma),\beta=\frac{-2}{N},C=\frac{c}{N\pi},D=\frac
{c}{N\pi m}.
\end{equation}
In this way we get a position space propagator which shows mass generation and
wave renormalization in all region. We can avoid infrared divergences in the
Euclid region.At large $N$ the function damps slowly with fixed $c,$where mass
changing effect is small for all range of $\left\vert x\right\vert $.For small
$N$ the function damps fast and the short distant part is dominant for mass
chaging effect.Short distance behaviour is determined by $D.$For long distance
we may treat finite $\mu$ and investigate the long distance behaviour of
$F.$In that case the $F$ contain only $\ln(\mu\left\vert x\right\vert )$ as
large $\mu\left\vert x\right\vert $ and others damp faster as $\ln
(\mu\left\vert x\right\vert )/\left\vert x\right\vert .$This indicate that the
mass generation is a short distance effect.In the fourier tranformation%
\begin{equation}
\rho(p)=\int_{0}^{\infty}\frac{\sin(p\left\vert x\right\vert )}{p\left\vert
x\right\vert }\left\vert x\right\vert ^{2}d\left\vert x\right\vert \frac
{\exp(-m\left\vert x\right\vert )}{4\pi\left\vert x\right\vert }\exp(F),
\end{equation}
first factor%
\[
\frac{\sin(p\left\vert x\right\vert )}{p\left\vert x\right\vert }%
\]
is dominant at small $\left\vert x\right\vert $ for both large and small $p.$

\section{\bigskip In momentum space}

\subsection{spectraly weighted propagator}

Now we turn to the fermion propagator.The momentum space propagator is given
by fourier transform%
\begin{align}
S_{F}(p)  &  =\int d^{3}x\exp(-ip\cdot x)S_{F}(x)\nonumber\\
&  =-\int d^{3}x\exp(-ip\cdot x)(i\gamma\cdot\partial+m)\frac{\exp
(-m\left\vert x\right\vert )}{4\pi\left\vert x\right\vert }\exp(F(x)).
\end{align}
where we have
\begin{equation}
F(x)=(c\left\vert x\right\vert )^{D-C\left\vert x\right\vert },D=\frac{c}{N\pi
m},C=\frac{c}{N\pi}.
\end{equation}
It is known that%
\begin{align}
&  \int d^{3}x\exp(-ip\cdot x)\frac{\exp(-m\left\vert x\right\vert )}%
{4\pi\left\vert x\right\vert }(c\left\vert x\right\vert )^{D}\nonumber\\
&  =c^{D}\frac{\Gamma(D+1)\sin((D+1)\arctan(\sqrt{-p^{2}}/m)}{\sqrt{-p^{2}%
}(p^{2}+m^{2})^{(D+1)/2}}\\
&  =\frac{1}{p^{2}+m^{2}}\text{ for }D=0,\nonumber\\
&  =\frac{2m}{(p^{2}+m^{2})^{2}}\text{ for }D=1.
\end{align}
for Euclidean momentum $p^{2}\leq0$.In Minkowski momentum $p^{2}\geq m^{2}$
above formula is continued to%
\begin{align}
\arctan\text{h}(z)  &  =\frac{1}{2}\ln(\frac{1+z}{1-z})=-i\arctan(iz),(0\leq
z^{2}\leq1)\\
\arctan\text{h}(z)  &  =\frac{1}{2}\ln(\frac{1+z}{z-1})\pm\frac{i\pi}{2},\\
\text{arccoth}(z)  &  =\frac{1}{2}\ln(\frac{1+z}{z-1}),(1\leq z^{2}).
\end{align}
From this definition
\begin{align}
\sinh(z)  &  =\frac{\exp(z)-\exp(-z)}{2}=-i\sin(iz),\nonumber\\
\sinh((D+1)\tanh^{-1}(z))  &  =\frac{-i}{2}[(\frac{1+z}{z-1})^{(D+1)/2}%
-(\frac{z-1}{1+z})^{(D+1)/2}],\nonumber\\
z  &  =\sqrt{p^{2}}/m.
\end{align}
The spectral function is a discontinuity in the upper half-plane of $z$
\begin{align}
\pi\rho(z)  &  =-\frac{c^{D}}{2}\operatorname{Im}\frac{\Gamma(D+1)}%
{m\left\vert z\right\vert (m^{2}-m^{2}z^{2})^{^{(D+1)/2}}}[(\frac{1+z}%
{1-z})^{(D+1)/2}-(\frac{1-z}{1+z})^{(D+1)/2}]\nonumber\\
&  =-\frac{c^{D}\Gamma(D+1)}{2mzm^{D+1}}\operatorname{Im}[(\frac{1}%
{1-z})^{D+1}\theta(z-1)-(\frac{1}{1+z})^{D+1}\theta(-(z+1)].
\end{align}
There is a problem to normalize the spectral function to $\delta(p^{2}-m^{2})$
in the weak coupling limit[5].Here we do not evaluate the explicit spectral
function.Principal part of the propagator in Minkowski space is continued to%
\begin{align}
S_{F}(p)  &  =\frac{(\gamma\cdot p+m)c^{D}\Gamma(D+1)\sinh((D+1)\text{arctanh}%
(\sqrt{p^{2}}/m)}{\sqrt{p^{2}}(m^{2}-p^{2})^{(D+1)/2}},(\left\vert \sqrt
{p^{2}}/m\right\vert \leq1)\\
&  =\frac{(\gamma\cdot p+m)c^{D}\Gamma(D+1)\sinh((D+1)\text{arccoth}%
(\sqrt{p^{2}}/m)}{\sqrt{p^{2}}(p^{2}-m^{2})^{(D+1)/2}},(\left\vert \sqrt
{p^{2}}/m\right\vert \geq1.\nonumber
\end{align}
To use above formula for $C\neq0$ case we use Laplace transform[7]
\begin{equation}
F(s)=\int_{0}^{\infty}d\left\vert x\right\vert \exp(-(s-m)\left\vert
x\right\vert )\left(  \mu\left\vert x\right\vert \right)  ^{-C\left\vert
x\right\vert }(s\geq0).
\end{equation}
This function shift the mass and we get the propagator
\begin{equation}
S_{F}(p)=(\gamma\cdot p+m)c^{D}\Gamma(D+1)\int_{0}^{\infty}F(s)ds\frac
{\sin((D+1)\arctan(\sqrt{-p^{2}}/(m-s))}{\sqrt{-p^{2}}(p^{2}+(m-s)^{2}%
)^{(D+1)/2}}.
\end{equation}
At $D=0$ and $1$ we see
\begin{equation}
S_{F}(p)=(\gamma\cdot p+m)\int_{0}^{\infty}F(s)ds\frac{1}{(p^{2}+(m-s)^{2}%
)},(D=0).
\end{equation}%
\begin{equation}
S_{F}(p)=(\gamma\cdot p+m)c\int_{0}^{\infty}F(s)ds\frac{2\left\vert
m-s\right\vert }{(p^{2}+(m-s)^{2})^{2}},(D=1).
\end{equation}
\qquad\ 

\section{Renormalization constant and order parameter}

Hereafter we consider the renormalization constant and bare mass in our
model.It is easy to evaluate the renormalization constant and bare mass by
define the renormalization
\begin{align}
\psi_{0}  &  =\sqrt{Z_{2}}\psi_{r},\overline{\psi}_{0}=\sqrt{Z_{2}}%
\overline{\psi}_{r},\nonumber\\
S_{F}^{0}  &  =Z_{2}S_{F},\frac{Z_{2}^{-1}}{\gamma\cdot p-m_{0}}=S_{F}(p),
\end{align}%
\begin{align}
Z_{2}^{-1}  &  =\int\rho_{1}(s)ds=\lim_{p\rightarrow\infty}\frac{1}{4p^{2}%
}tr(\gamma\cdot pS_{F}(p))\nonumber\\
&  =\Gamma(D+1)c^{D}\lim_{p\rightarrow\infty}\int_{0}^{\infty}G(s)ds\frac
{\sqrt{p^{2}}\sin((D+1)\arctan(\sqrt{p^{2}}/(m-s))}{(p^{2}+(m-s)^{2}%
)^{(1+D)/2}}\nonumber\\
&  \rightarrow\sin(\frac{(D+1)\pi}{2})c^{D}\lim_{p\rightarrow\infty}\frac
{1}{p^{D}}=0.
\end{align}%
\begin{equation}
m_{0}Z_{2}^{-1}=m\int\rho_{2}(s)ds=\lim_{p\rightarrow\infty}\frac{1}%
{4}tr(p^{2}S_{F}(p))\rightarrow0.
\end{equation}
There is no pole and it shows the confinement for $D>0$.Order parameter
$\left\langle \overline{\psi}\psi\right\rangle $ is given as the integral of
the scalar part of the propagator in momentum space
\begin{equation}
\left\langle \overline{\psi\psi}\right\rangle =-TrS_{F}(x),
\end{equation}
\begin{align}
\left\langle \overline{\psi}\psi\right\rangle  &  =-2\int_{0}^{\infty}%
\frac{p^{2}d\sqrt{p^{2}}}{2\pi^{2}}\frac{\Gamma(D+1)c^{D}}{2\sqrt{-p^{2}}%
}\nonumber\\
&  \times\int_{0}^{\infty}dsF(s)\frac{\sin((D+1)\arctan(\sqrt{p^{2}}%
/(m-s))}{\sqrt{((m-s)^{2}+p^{2})^{D+1}}},(D=c/N\pi m).
\end{align}
For $D=c/N\pi m\geq1$ vacuum expectation value is finite.This condition is
independent of the bare mass.We cannot apriori determine the value $D,$since
$m$ is a physical mass$.m$ and $\Sigma(0)$ is not the same quantity but we
assume they have the same order of maginitude.In numerical analysis of
Dyson-Schwinger equation $\Sigma(0)$ damps fast with $N$ and is seen to vanish
at $N=3.$If we assume $m=O(c/N)$ in the case of vanishing bare mass $m_{0}%
=0$,$D$ becomes $O(1).$In our approximation if we set $m$ equals to the second
order value in the Landau gauge,we get $D\simeq1.08$ which is very close
to$1$.For $D=1$ we have a similar solution of the propagator at short distance
which is known by the analysis of D-S.In the analysis of Gap equation,zero
momentum mass $\Sigma(0)$ and a small critical number of flavours have been
shown[1,2,3],in which same approximation was done because the vacuum
polarization governs the photon propagator at low energy.The critical number
of flavour $N_{c}$ is a consequence of the approximation for the infrared
dynamics as in quenched QED$_{4}$ where ultraviolet rigion is non trivial and
is not easy to find numerically.In the intermeadiate value of the coupling we
solved the coupled Dyson-Schwinger equation numerically and found that the
$\Sigma(0)$ which is,$O(e^{2}/4\pi)$ at $N=1$,the same order of magnitude as
the quenched Landau gauge[15].

\section{Coulomb-energy and self-energy}

In this section we study the origin of confinement.First we see the difference
between short and long distance cases.The coefficent of $\ln(c\left\vert
x\right\vert ),c\left\vert x\right\vert \ln(c\left\vert x\right\vert )$ in two
cases of $F.$If we expand the $D_{F}(x)$ in $c\left\vert x\right\vert ,1/$
$c\left\vert x\right\vert ,$for the latter case%
\begin{equation}
D_{F}(c\left\vert x\right\vert )\simeq\frac{4c^{2}}{N\pi^{2}}\frac{1}%
{c^{2}x^{2}},(c\left\vert x\right\vert \gg1)
\end{equation}
we easily obtain%
\begin{equation}
F=\frac{8}{N\pi^{2}}\ln(c\left\vert x\right\vert ),(c\left\vert x\right\vert
\gg1).
\end{equation}%
\begin{align}
F  &  =\frac{(d-2)}{N}\ln(\frac{\mu}{c})+\frac{\gamma}{N\pi m}+\frac{c}{N\pi
m}\ln(c\left\vert x\right\vert )\nonumber\\
&  -\frac{c}{N\pi}\left\vert x\right\vert \ln(c\left\vert x\right\vert
)-\frac{c}{2N\pi}\left\vert x\right\vert (d+3-2\gamma)\nonumber\\
,(c\left\vert x\right\vert  &  \ll1).
\end{align}
Here%
\begin{align}
D  &  =\frac{c}{N\pi m},(c\left\vert x\right\vert \ll1),\nonumber\\
&  =\frac{8}{N\pi^{2}},(c\left\vert x\right\vert \gg1),\\
E  &  =\frac{1}{N\pi},(c\left\vert x\right\vert \ll1)\nonumber\\
&  =0,(c\left\vert x\right\vert \gg1).
\end{align}
are both the coefficents of the Coulomb energy and self-energy which we will
see below.Our approximation to the spectral function $T_{1}\overline{T}_{1}$
is an four particle scattering amplitude with one photon exchange.In the
evaluation of $F$ we see that $D$ term comes from the vertex by LSZ which has
an infrared singularity%
\begin{equation}
\frac{1}{\gamma\cdot(p+k)-m+i\epsilon}\gamma_{\mu}\epsilon^{\mu}%
(k,\lambda)U(p,s)\rightarrow\frac{\gamma\cdot(p+k)+m}{2p\cdot k}\gamma_{\mu
}\epsilon^{\mu}(k,\lambda).
\end{equation}
In this way the vertex is enhanced near the on-shell as $1/k$ in contrast with
the usual on-shell matrix element
\begin{equation}
\overline{U}(p+k)\gamma_{\mu}U(p)\epsilon^{\mu}(k)
\end{equation}
$E$ term comes from $1/(p\cdot k)^{2}$ and $1/(p\cdot k).$In qenched case it
comes from $1/(p\cdot k)^{2}.$Let us imagine the imaginary part of the fermion
propagator.When $p^{2}\geq m^{2}$ parent fermion can decay into fermion and
photon,this process gives a Coulomb energy.Namely%
\begin{align}
\int_{m}^{\infty}ds  &  \int d^{3}y\frac{\exp(-s\left\vert y\right\vert
)}{4\pi\left\vert y\right\vert }\frac{\exp(-s\left\vert x-y\right\vert )}%
{4\pi\left\vert x-y\right\vert }F(x-y)\nonumber\\
&  =\int d^{3}y\rho(x-y)\rho(y)(D\ln(c\left\vert x-y\right\vert )-Cc\left\vert
x-y\right\vert \ln(c\left\vert x-y\right\vert )).
\end{align}
When $p^{2}\leq m^{2}$ parent cannot dacay,this one gives a fermion
self-energy.%
\begin{equation}
\delta^{(3)}(0)\frac{\exp(-m\left\vert x\right\vert )}{4\pi\left\vert
x\right\vert }F\simeq V(\frac{D\ln(c\left\vert x\right\vert )}{\left\vert
x\right\vert }-Cc\left\vert x\right\vert \ln(c\left\vert x\right\vert )).
\end{equation}
Usually these are summed as leading infrared divergences.It may be clear if we
consider the potential energy for two charged particle with modified Coulomb
interaction and seek the corresponding terms in $S_{0}F$.At short distance wth
bare photon we have%
\begin{align}
V_{C}(\left\vert x\right\vert )  &  =\int\frac{d^{2}k}{(2\pi)^{2}}\exp(ik\cdot
x)\frac{1}{k^{2}(p\cdot k)}\nonumber\\
&  =\frac{1}{4\pi}\int_{0}^{\infty}d\alpha K_{0}(x+ap,\mu)\propto\left\vert
x\right\vert \ln(c\left\vert x\right\vert ),(c\left\vert x\right\vert \ll1).
\end{align}
At long distance with dressed photon we have%
\begin{equation}
V_{C}(\left\vert x\right\vert )=\int\frac{d^{2}k}{(2\pi)^{2}}\exp(ik\cdot
x)\frac{m^{2}}{ck(p\cdot k)}\propto\frac{1}{c\left\vert x\right\vert
}(c\left\vert x\right\vert \gg1).
\end{equation}
In the same way we obtain the contribution from another term which is singular
as $1/(p\cdot k)^{2}$%
\begin{align*}
V_{S}(\left\vert x\right\vert )  &  =\int\frac{d^{2}k}{(2\pi)^{2}}\exp(ik\cdot
x)\frac{m^{2}}{k^{2}(p\cdot k)^{2}}\\
&  =\frac{1}{4\pi}\int\alpha d\alpha K_{0}(x+ap,\mu)\propto\left\vert
x\right\vert ^{2}\ln(c\left\vert x\right\vert )(c\left\vert x\right\vert
\ll1),
\end{align*}%
\begin{equation}
V_{S}(\left\vert x\right\vert )=\int\frac{d^{2}k}{(2\pi)^{2}}\exp(ik\cdot
x)\frac{1}{ck(p\cdot k)^{2}}\propto\ln(c\left\vert x\right\vert )(c\left\vert
x\right\vert \gg1),
\end{equation}
which is not familiar to us.However this term drives confinement at long
distance as in the queched case.In many cases we assume the absence of mass
changing effects.Of course the non-relativistic approximation is correct in
that sense.But the $m^{2}/(p\cdot k)^{2}$ has also an infrared singular
contribution to the self-energy.This one create the position dependent mass as
$M(x)=c\ln(c\left\vert x\right\vert ).$The coefficents $D$,$E$ are gauge
invariant provided the photon couples to conserved currents.

\section{Summary}

We evaluate the fermion propagator in three dimensional QED with dressed
photon by method of spectral function.Non perturbative effects are included by
resummation of the infinite numbers of rainbow type diagrams with physical
mass.In the evaluation of lowest order matrix element for fermion spectral
function we obtain finite mass shift,Coulomb energy and position dependent
mass,which is silmilar to the analysis of D-S equation excepts for the wave
function renormalization.Including vacuum polarization we find the same
structure.Above some coupling constant order parameter $\left\langle
\overline{\psi}\psi\right\rangle $ is finite,which is independent of the
symmetry which forbids finite bare mass.The arguments of confinement are given
usually for the force between charged paricle.In our approximation these force
sets the renormalization constant $Z_{2}^{-1}=0$ for arbitrary coupling.If we
assume the magnitude of $m$ is generated by the second order in $e$ in the
Landau gauge,our results is consistent with numerical analysis of coupled
Dyson-Schwinger equation [15].

\bigskip

\section{References\newline}

\noindent\lbrack1]T.Appelquist,D.Nash,Critical Behaviour in (2+1)-Dimensional
QED, Phys.Rev.Lett.\textbf{60 }(1988)2575.\newline%
[2]T.Appelquist,L.C.R.Wijewarhana,Critical Behaviour in (2+1)-Dimensional QCD,
\ Phys.Rev.Lett.\textbf{64},721(1990).\newline%
[3]T.Appelquist,L.C.R.Wijewarhana,Phase Structure of Non-Compact QED3 and the
Abelian Higgs Model,hep-ph/0403250.\newline[4]Y.Hoshino,T.Matsuyama,Dynamical
parity violation in QED in three-dimensions with a two component massless
fermion,Phys.lett.\textbf{B} \textbf{222}(1989)493.\newline%
[5]R.Jackiw,L.Soloviev,Low-energy theorem approach to single-particle
singularities in the presence of massless bosons,Phys.Rev.\textbf{137}%
.3(1968)1485.\newline[6]A.B.Waites,R.Delbourgo,Non pertubative behaviour in
three-dimensional QED,Int.J.Mod.Phys.\textbf{A7}(1992)6857.\newline[7]Yuichi
Hoshino,Mass singularity and confining property in QED3,\textbf{JHEP0409}%
:048,2004.\newline[8]S.Deser,R.Jackiw,S.Templeton,Topologically massive gauge
theory,Ann,Phys.(NY)\textbf{140 }(1982)372.\newline%
[9]T.Appelquist,M.Bowick,D.Karabari and L.C.R.Wijewardhana,Spontaneous
Breaking of Parity in (2+1)-dimensional QED,Phys.Rev.\textbf{D33}%
(1986)3774.\newline[10]K.Nishijima,\textbf{Fields and Particles}%
,W.A.BENJAMIN,INC(1969).\newline[11]C.Itzykson,J.B.Zuber,Quantum field
theory,McGRAW-HILL.\newline[12]L.S.Brown,Quantum field theory,Cambridege
University Press,1992.\newline[13]J D.Bjorken,S D.Drell,Relativistic Quantum
Mechanics,McGraw-Hill Book Company.\newline[14]Yuichi Hoshino,in
preparation.\newline[15]Yuichi Hoshino,Toyoki Matsuyama and Chikage
Habe,Fermion mass generation in in QED in three-dimensions,in the proceedings
of the workshop on Dynamical symmetry Breaking,Nagoya,1989:0227-232.\newline

\end{document}